\documentclass[conference]{IEEEtran}
\IEEEoverridecommandlockouts

\usepackage{amsmath, amssymb, amsfonts}
\usepackage{graphicx}
\usepackage{booktabs}
\usepackage{cite}
\usepackage{float}
\usepackage{hyperref}
\usepackage{siunitx}
\usepackage{caption}
\usepackage{subcaption}
\usepackage{multirow}
\usepackage{listings}  

\begin{document}

\title{
A Hybrid Jump-Diffusion Model for Coherent Optical Control of Quantum Emitters in hBN
}

\author{
Saifian Farooq Bhat$^{1}$,
Michael K. Koch$^{2}$,
Sachin Negi$^{1}$,
Alexander Kubanek$^{2}$,
Vibhav Bharadwaj$^{1,*}$\\[0.5em]
\small
$^{1}$Department of Physics, Indian Institute of Technology Guwahati, India\\
$^{2}$Institute for Quantum Optics, Ulm University, Baden-Württemberg, Germany\\
$^{*}$Corresponding author\\
Email: s.bhat@iitg.ac.in, vibhav9@iitg.ac.in
}

\maketitle

\begin{abstract}
Hexagonal boron nitride (hBN) has emerged as a promising two-dimensional host for stable single-photon emission owing to its wide bandgap, high photostability, and compatibility with nanophotonic integration. We present a simulation-based study of temperature-dependent spectral dynamics and optical coherence in a mechanically decoupled quantum emitter in hBN. Employing a hybrid stochastic framework that combines Ornstein-Uhlenbeck detuning fluctuations with temperature-dependent, Gaussian-distributed discrete frequency jumps, motivated by experimentally observed spectral diffusion and blinking, we reproduce the measured evolution of inhomogeneous linewidth broadening and the progressive degradation of photon coherence across the relevant cryogenic range (5-30~K). The model captures phonon-related spectral diffusion with a cubic temperature dependence and the onset of jump-like spectral instabilities at higher temperatures. By calibrating the hybrid diffusion, jump parameters to the experimentally measured full width at half maximum (FWHM) of the emission line and analyzing the second-order correlation function $g^{(2)}(\tau)$ under resonant driving, we establish a unified phenomenological description that links stochastic detuning dynamics to the decay of optical coherence in a resonantly driven emitter. Analysis of $g^{(2)}(\tau)$ under resonant driving reveals an additional dephasing rate $\gamma_{\mathrm{sd+j}}$ that rises monotonically with temperature and drive strength, leading to a predicted critical crossover to overdamped dynamics at $T_{\mathrm{crit}} \approx 25.91$~K. This hybrid framework provides a quantitative connection between accessible spectroscopic observables and the dominant noise mechanisms limiting coherent optical control in mechanically decoupled quantum emitters, exemplified in hBN and generalizable to similar emitters in other materials.
\end{abstract}

\section{Introduction}
hBN possesses a wide bandgap of approximately $5.9~\text{eV}$~\cite{Cassabois2016,Tran2016}, supporting deep-ultraviolet to near-infrared emission arising from defect-associated electronic transitions~\cite{Jungwirth2016,Kianinia2017,Mendelson2021}. Owing to their brightness, polarization stability, and integrability, hBN quantum emitters have emerged as promising alternatives to semiconductor quantum dots for stable single-photon sources, provided that appropriate device engineering (encapsulation, substrate choice, or mechanical decoupling) is applied to mitigate spectral diffusion and sample variability~\cite{Aharonovich2016,Atature2018,Aharonovich2022hBN}.
In mechanically decoupled quantum emitters in hBN, reduced coupling to substrate-induced strain and charge noise has been shown to result in enhanced spectral stability and narrow optical resonances at cryogenic temperatures, including near 660~nm in representative system~\cite{Koch2024}. This mechanical isolation enables coherent optical control and exposes intrinsic phonon-limited coherence that is otherwise masked in substrate-bound samples.
Despite these advantages, the optical coherence of hBN defect centers remains strongly influenced by phononic and electrostatic fluctuations. Temperature-dependent spectroscopy has shown that spectral diffusion, phonon coupling, and charge-induced dephasing can jointly broaden the emission linewidth and degrade coherent control of emitters in hBN~\cite{Dietrich2020,Exarhos2017,Gottscholl2021}. For specific classes of emitters~\cite{Koch2024}, both the zero-phonon line (ZPL) shift and linewidth have been reported to follow a cubic temperature dependence, consistent with deformation-potential coupling to low-energy acoustic phonons~\cite{Grange2017}. However, alternative temperature scalings have been reported for other hBN emitters, reflecting defect-specific phonon-coupling mechanisms and local environments~\cite{Optica2021}. Photoluminescence excitation spectroscopy further reveals that, in such emitters, an energy gap between the ZPL and the first acoustic mode of the phonon sideband can close with increasing temperature, leading to a rapid increase in dephasing once phonon protection is lost~\cite{Koch2024}.
Beyond homogeneous phonon-induced broadening, hBN emitters also exhibit significant inhomogeneous broadening caused by slow spectral diffusion. Local charge reconfigurations, strain fluctuations, and defect rearrangements induce stochastic frequency shifts that broaden time-averaged spectra beyond the lifetime-limited linewidth, even at cryogenic temperatures.
For the mechanically decoupled quantum emitter in hBN that serves as the experimental reference throughout this work (Ref.~\cite{Koch2024}), the measured linewidth follows
\begin{equation}
\Gamma(T) = A + B T^3,
\label{eq:fwhm_T}
\end{equation}
with $A = 1.01~\mathrm{GHz}$ and $B = 3.77\times10^{-5}~\mathrm{GHz/K}^3$. The constant term reflects radiative lifetime broadening together with residual quasi-static spectral diffusion, while the cubic term arises from acoustic-phonon-mediated dephasing. Mechanical decoupling suppresses the temperature-independent inhomogeneous component, but at higher temperatures, thermally activated charge motion and lattice fluctuations reintroduce spectral wandering and discrete frequency jumps that dominate the emitter’s dynamics (see Supplementary Information, Sec.~III).
Coherent optical control--including Rabi oscillations and $\pi$-pulse excitation--has been demonstrated up to $\sim 30~\mathrm{K}$ in mechanically isolated hBN~\cite{Koch2024,Branny2017,Kianinia2021}. At elevated temperatures, however, activated lattice noise and charge fluctuations lead to a transition from coherent to overdamped behavior governed by both continuous spectral wandering and abrupt jump-like instabilities~\cite{Bourassa2020,Kumar2022}. Reports of near-Fourier-limited coherence approaching room temperature in certain hBN-based emitter platforms, including suspended and nanophotonic device geometries, highlight the strong sensitivity of coherence preservation to fabrication quality, suspension geometry, and the local nano-environment~\cite{Dietrich2020,Zeng2020,Kianinia2021,Koch2024}.
In this work, we develop a hybrid stochastic model inspired by recent theoretical treatments of spectral diffusion in solid-state emitters~\cite{Delteil2024,Gerard2025}, combining Ornstein-Uhlenbeck diffusion with a discrete Gaussian jump process to capture both continuous and abrupt detuning fluctuations (see Supplementary Information, Sec.~I for numerical implementation and Sec.~III for the physical motivation of jumps). Using experimentally extracted temperature-dependent parameters, the model reproduces the measured linewidth evolution, the power-dependent damping of optical Rabi oscillations, and the temperature-dependent behavior of the second-order correlation function $g^{(2)}(\tau)$ (see Supplementary Information, Fig.~S2 and Sec.~VII). This unified framework links phonon-driven diffusion with jump-like spectral instabilities and enables quantitative simulation of coherence dynamics in mechanically decoupled hBN emitters.

\section{Theoretical Model and Simulation Framework}
The spectral dynamics of a single quantum emitter in hBN are modeled using a hybrid stochastic approach that captures both continuous and discrete detuning fluctuations. This framework is motivated by recent theoretical treatments of spectral diffusion in solid-state emitters~\cite{Delteil2024,Gerard2025} and by foundational studies of Gaussian and telegraph-like noise processes in semiconductor defects. The model combines Ornstein-Uhlenbeck (OU) diffusion, representing phonon-induced dephasing, with a Gaussian Random Jump (GRJ) process describing abrupt frequency shifts associated with charge motion, defect rearrangements, or local strain reconfigurations (see Supplementary Information, Sec.~III).
In the discrete form implemented numerically, the instantaneous detuning $\omega_i$ evolves according to
\begin{equation}
\omega_{i+1} = \omega_i - \frac{\big(\omega_i - \omega_0\big)}{\tau_{\mathrm{sd}}}\,dt
+ S\sqrt{\frac{1}{2\tau_{\mathrm{sd}}}}\,\xi_i\sqrt{dt} + J_i
\label{eq:update_rule}
\end{equation}
where $\omega_0$ is the mean transition frequency, $\tau_{\mathrm{sd}}$ is the spectral diffusion correlation time, and $S$ determines the strength of the diffusion term. The discrete noise $\xi_i$ is drawn from a standard normal distribution $\mathcal{N}(0,1)$. The final term $J_i$ models spectral jumps: jumps occur with probability $P_J = \lambda_J\,dt$, and each amplitude is sampled from a Gaussian distribution $\mathcal{N}(0,\sigma_J^2)$ (see Supplementary Information, Sec.~I for full implementation details).
Taking the continuous-time limit yields the stochastic differential equation
\begin{equation}
d\omega(t) = -\frac{\omega(t) - \omega_0}{\tau_{\mathrm{sd}}}\,dt
+ S\sqrt{\frac{1}{2\tau_{\mathrm{sd}}}}\,dW_t + J(t)
\label{eq:continuous_ou_jump}
\end{equation}
where $dW_t$ is the Wiener increment and $J(t)$ is a compound Poisson process with rate $\lambda_J$ and Gaussian jump amplitudes.
This hybrid diffusion-jump dynamics produces an effective inhomogeneously broadened emission line whose full width at half maximum (FWHM) is approximated by
\begin{equation}
\Gamma_{\mathrm{sim}} =
2\sqrt{2\ln 2}\,
\sqrt{\frac{S^2}{2} +
      \frac{\lambda_J\sigma_J^2 \tau_{\mathrm{sd}}}{2}}
\label{eq:gamma_sim}
\end{equation}
The exact analytical stationary variance leading to Eq.~\eqref{eq:gamma_sim} is derived in Supplementary Information, Eq.~(9).
To ensure physical consistency, $\Gamma_{\mathrm{sim}}$ is calibrated against the empirical broadening law
\begin{equation}
\Gamma_{\exp}(T) = 1.01 + 3.77 \times 10^{-5} T^3~\text{GHz},
\label{eq:gamma_exp}
\end{equation}
as measured in the mechanically decoupled hBN emitter of Koch \textit{et al.}~\cite{Koch2024} (see Supplementary Information, Sec.~VI and Table~I for the complete two-stage calibration protocol using physically motivated temperature-dependent multipliers).
The parameters $(S, \lambda_J, \sigma_J)$ are treated as temperature dependent and increase monotonically with $T$. For each temperature, Eq.~\eqref{eq:update_rule} is integrated numerically using the Euler-Maruyama method over a 10~ns window, and the resulting detuning histograms are fitted to extract $\Gamma_{\mathrm{sim}}(T)$ (see Supplementary Information, Sec.~V).

\section{Simulation Results}
The calibrated jump-diffusion model reproduces the experimentally observed temperature dependence of the emission linewidth of the specific mechanically decoupled hBN quantum emitter studied in Ref.~\cite{Koch2024}. For each temperature, the stochastic parameters $(S, \lambda_J, \sigma_J)$ were obtained through a constrained calibration in which their values were adjusted monotonically to match the experimentally measured inhomogeneous broadening $\Gamma_{\mathrm{exp}}(T)=A+BT^{3}$ with an accuracy better than $0.02~\mathrm{GHz}$ (see Supplementary Information, Sec.~VI). The calibrated parameters provide qualitative insight into the relative contributions of continuous diffusion and jump-like detuning processes, and the resulting agreement between simulation and experiment is shown in Fig.~1, where the simulated linewidths reproduce the cubic, phonon-limited trend across the experimentally relevant temperature range from 5 to 30~K.

\begin{figure}[H]
    \centering
    \includegraphics[width=1\linewidth]{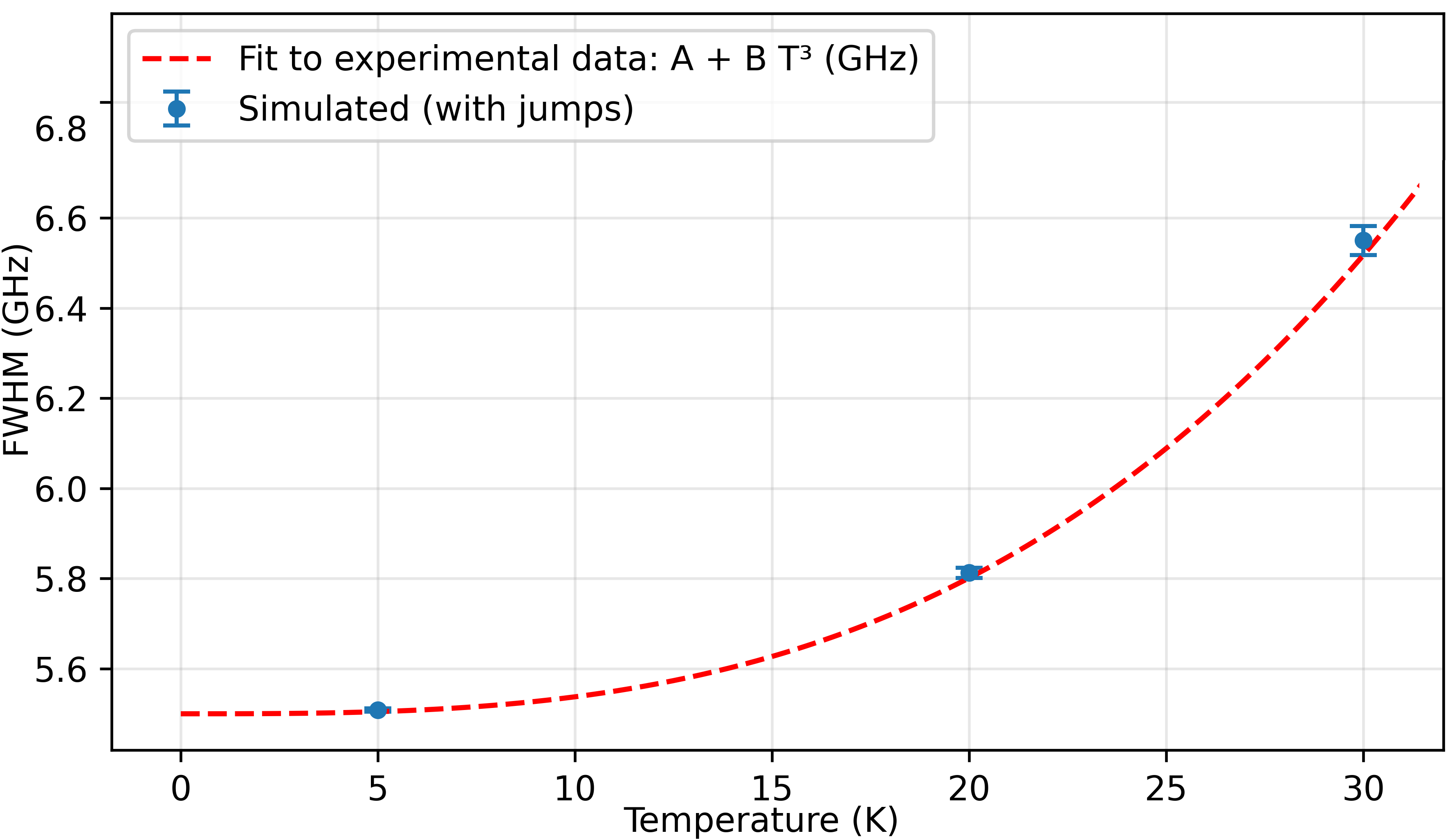}
    \caption{\footnotesize Simulated and experimental inhomogeneous linewidth broadening as a function of temperature. The simulated FWHM values (blue) closely follow the empirical $A + BT^3$ trend (red dashed line), confirming the expected phonon-limited broadening behavior.}
\label{fwhm_vs_T}
\end{figure}

\begin{figure}[H]
    \centering
    \includegraphics[width=1.0\linewidth]{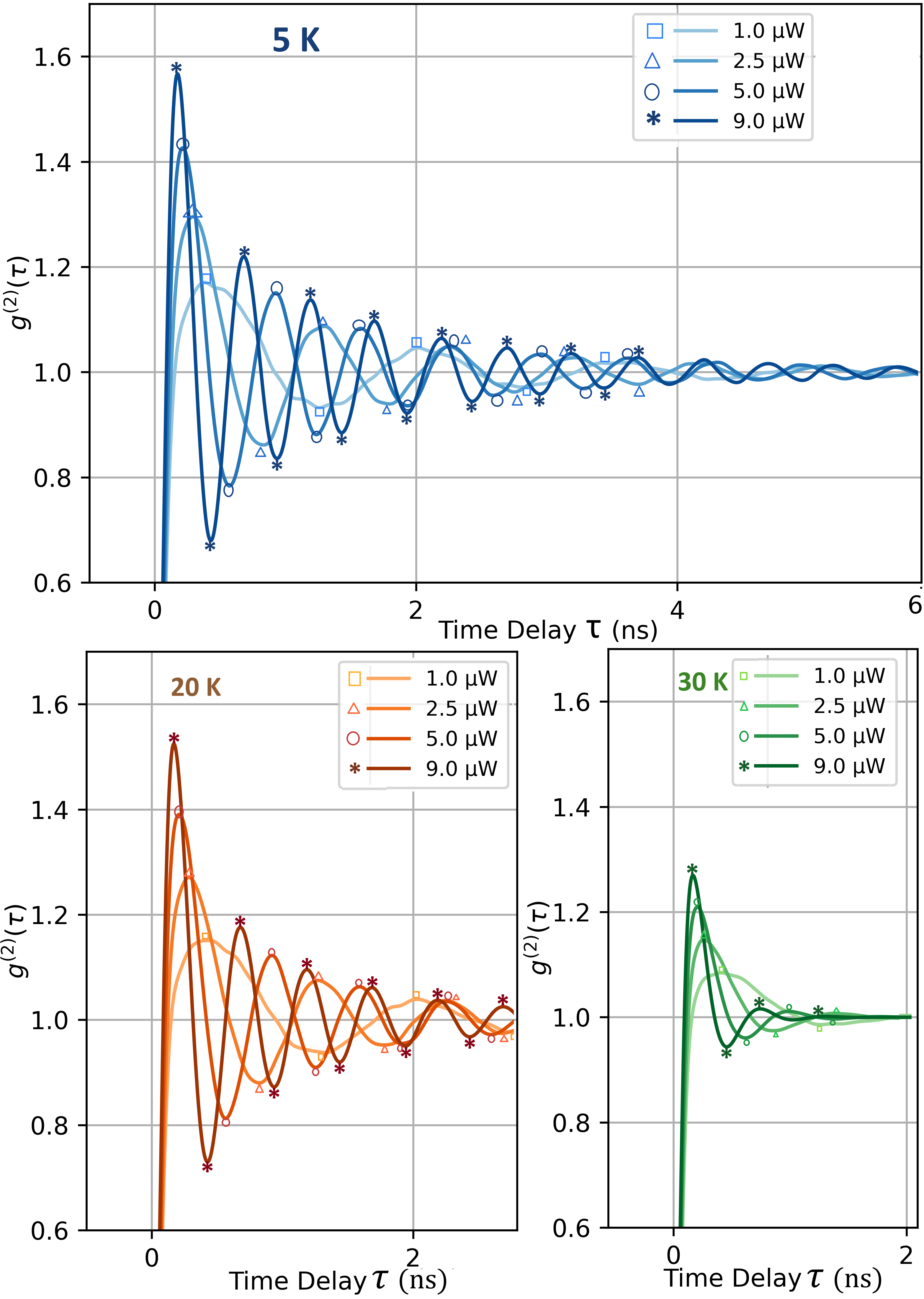}
    \caption{\footnotesize (a-c) Simulated second-order correlation functions $g^{(2)}(\tau)$ for the hBN quantum emitter studied in Ref.~\cite{Koch2024} at 5~K, 20~K, and 30~K, respectively.
    Increasing excitation power increases the Rabi frequency ($\Omega_R \propto \sqrt{P}$), which manifests in $g^{(2)}(\tau)$ as more pronounced oscillatory shoulders and a higher oscillation frequency at short delay times. At elevated temperatures, these oscillatory features are progressively damped due to enhanced spectral diffusion and thermally activated frequency jumps, leading to suppression of coherent Rabi dynamics (see Supplementary Information, Fig.~S2 for parameter dependence).}
    \label{g2_5K}
\end{figure}

\begin{figure}[H]
    \centering
    \includegraphics[width=1.0\linewidth]{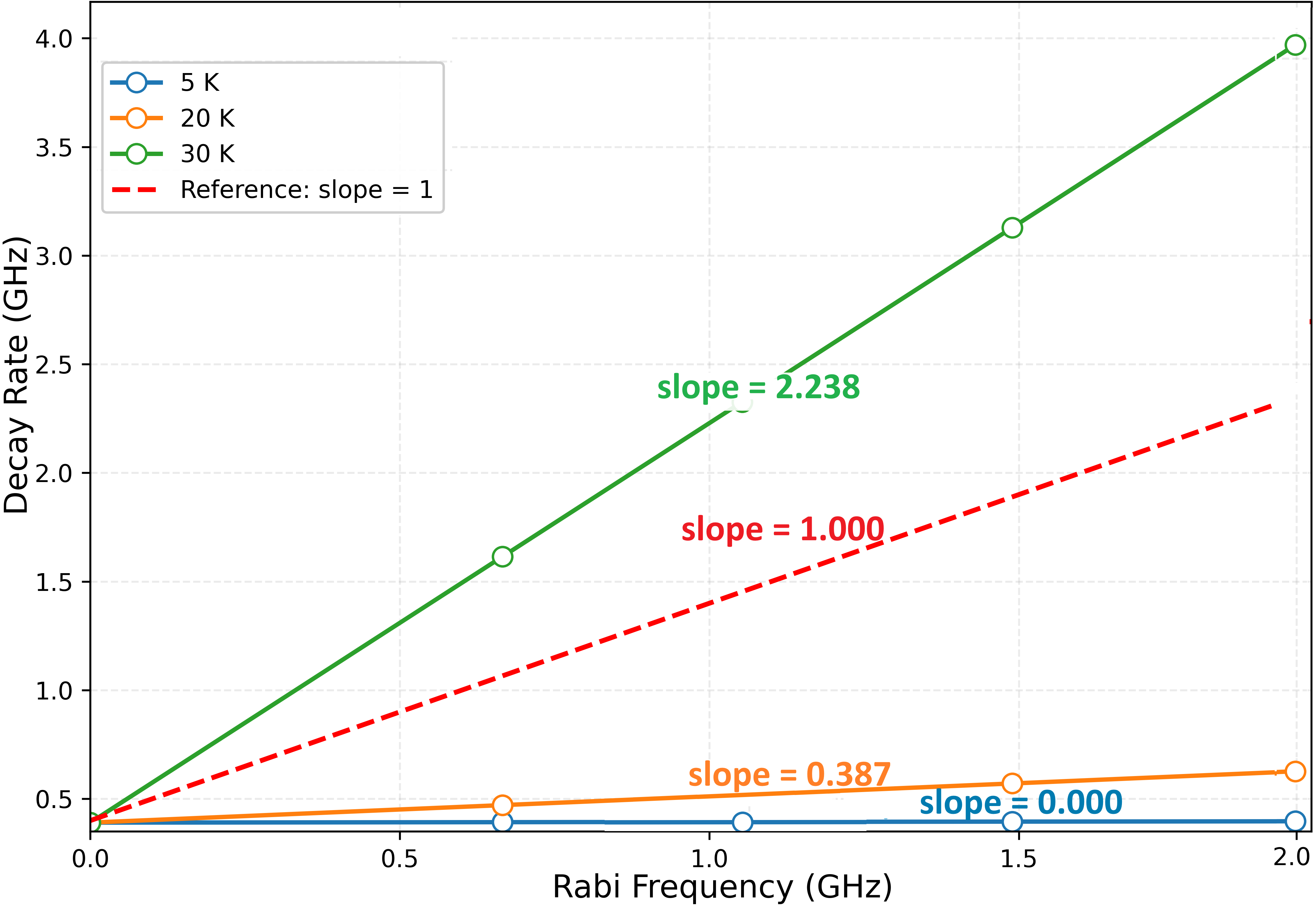}
    \caption{\footnotesize Decay rate extracted from simulated $g^{(2)}(\tau)$ traces as a function of Rabi frequency for 5~K, 20~K, and 30~K.
The dashed red line marks the $\Gamma = \Omega_R$ boundary separating coherent and overdamped regimes.
The increase in slope with temperature reflects the monotonic rise of $\gamma_{\mathrm{sd+j}}$, which captures spectral-diffusion- and jump-induced inhomogeneous broadening.}
\label{alpha_vs_rabi}
\end{figure}

The second-order photon correlation function $g^{(2)}(\tau)$ provides a sensitive probe of coherence and spectral dynamics in resonantly driven solid-state emitters. Its oscillatory behavior and damping are shown in Fig.~\ref{g2_5K}. The correlation function reflects the combined influence of population relaxation, pure dephasing, and stochastic detuning noise. Following the theoretical treatments in Refs.~\cite{Delteil2024,Gerard2025}, the driven two-level system under stochastic frequency fluctuations is described by an effective correlation function of the form
\begin{equation}
\begin{split}
g^{(2)}(\tau) &= 1 - \exp\!\left[-\frac{1}{2}\left(\frac{1}{T_1} + \frac{1}{T_2} + \gamma_{\mathrm{sd+j}}\right)|\tau|\right]\\
&\times \Bigg[\cos(\Omega_{\mathrm{eff}}\tau) +
\frac{\left(\tfrac{1}{T_1} + \tfrac{1}{T_2}\right)}{2\Omega_{\mathrm{eff}}}\sin(\Omega_{\mathrm{eff}}\tau)\Bigg],
\end{split}
\label{eq:g2_final}
\end{equation}
where $T_1$ and $T_2$ are the radiative lifetime and intrinsic coherence time, respectively, $\Omega_{\mathrm{eff}}$ is the effective Rabi frequency, and $\gamma_{\mathrm{sd+j}}$ represents the additional dephasing induced by spectral diffusion and discrete spectral jumps (see Supplementary Information, Sec.~VII for the derivation). As shown in Fig.~\ref{alpha_vs_rabi}, $\gamma_{\mathrm{sd+j}}$ increases monotonically with both temperature and Rabi frequency $\Omega_R$. This expression captures both the oscillatory behavior at low temperatures and the transition to an overdamped regime as stochastic detuning noise grows (see Supplementary Information, Fig.~S1 for the pure OU limit and Fig.~S2 for the hybrid-parameter study).

The effective decay rate extracted from the simulated photon-correlation traces provides a direct quantitative measure of how spectral diffusion and jump processes modify coherent optical dynamics. The additional dephasing term $\gamma_{\mathrm{sd+j}}$ appearing in Eq.~\eqref{eq:g2_final} encapsulates the combined influence of continuous OU-type spectral diffusion and discrete Gaussian jumps--precisely the same mechanisms responsible for the inhomogeneous broadening discussed earlier in the linewidth analysis. As shown in Fig.~\ref{alpha_vs_rabi}, $\gamma_{\mathrm{sd+j}}$ is negligible at 5~K, resulting in decay rates that remain nearly independent of the Rabi frequency. At 20~K, $\gamma_{\mathrm{sd+j}}$ increases from a negligible value and develops a weak but measurable linear dependence on the Rabi frequency $\Omega_R$, coinciding with a fivefold increase in the calibrated jump amplitude $\sigma_J$, while the diffusion strength $S$ and jump rate $\lambda_J$ remain unchanged. At 30~K, $\gamma_{\mathrm{sd+j}}$ increases further as the jump-induced detuning remains strong and the diffusion strength $S$ rises by approximately 10\%, leading to a steep increase in the decay rate that surpasses the $\Gamma=\Omega_R$ boundary. This marks the transition to an overdamped regime ($\Gamma=2\times\Omega_R$), in which noise-induced detuning dominates over coherent Rabi cycling. The resulting monotonic, temperature-driven enhancement of $\gamma_{\mathrm{sd+j}}$ establishes a direct link between inhomogeneous broadening, set by the combined action of spectral diffusion and jump processes. The observed suppression of coherent dynamics under resonant driving (see Supplementary Information, Fig.~S3, which demonstrates that only the hybrid model with discrete jumps reproduces the experimentally observed compact, near-Gaussian detuning distributions at elevated temperatures).

\section{Conclusion}
Our hybrid jump-diffusion model reproduces the experimentally observed temperature and power-dependent coherence degradation of the mechanically decoupled hBN quantum emitter studied in Ref.~\cite{Koch2024}, using only the measured $T^{3}$ inhomogeneous broadening law as input~\cite{Koch2024}. The framework captures the interplay between continuous spectral diffusion and discrete jump-like detuning processes responsible for coherence loss under resonant driving. While the numerical values of the calibrated parameters are specific to this emitter, the modeling approach can be extended to mechanically decoupled hBN quantum emitters exhibiting similar temperature-dependent inhomogeneous broadening. Full numerical implementation, parameter scans, calibration strategy, analytical derivations, and validation are provided in the Supplementary Information (Secs.~I-VII, Eqs.~(1)-(9), Figs.~S1-S3, and Table~I).
\textbf{Inversion of Calibration and Extraction of Critical Dephasing Temperature}\\
Most importantly, the model identifies the fundamental crossover where the effective dephasing rate equals the Rabi frequency (slope = 1). This crossover marks the transition from a coherently driven regime to an overdamped regime, in which stochastic detuning fluctuations dominate over coherent Rabi cycling. Beyond this crossover, reliable coherent control--such as the implementation of well-defined $\pi/2$ pulses within a single coherence time--becomes impractical due to rapid noise-induced dephasing. For the mechanically decoupled class of hBN emitters studied in Ref.~\cite{Koch2024}, this limit occurs at
\begin{equation}
T_{\mathrm{crit}} \approx 25.91~\text{K}.
\end{equation}
This temperature was obtained by identifying the diffusion parameters for which the slope equals unity and mapping the resulting simulated linewidth to temperature using the calibrated relation in Eq.~\eqref{eq:fwhm_T}.
Above this temperature, resonant optical control rapidly becomes incoherent regardless of excitation power. Crucially, the framework is fully general: it applies to \emph{any} solid-state quantum emitter exhibiting a cubic temperature dependence of the inhomogeneous linewidth, regardless of the material platform.
With just three experimental temperature points providing the empirical law $\Gamma(T) = A + B T^3$, the model performs inverse calibration to extract the full temperature evolution of the microscopic noise parameters as diffusion strength $S(T)$, diffusion correlation time $\tau_{\mathrm{sd}}(T)$, jump rate $\lambda_J(T)$, and jump amplitude $\sigma_J(T)$ and thereby predicts the critical dephasing temperature $T_{\mathrm{crit}}$ at which the stochastic dephasing rate equals the Rabi frequency (slope = 1), marking the ultimate limit of coherent optical control (see Supplementary Information, Sec.~VI and Table~I).
This predictive capability establishes a direct, quantitative link between experimentally accessible spectroscopic observables and the microscopic noise processes governing optical coherence. By disentangling the respective contributions of continuous spectral diffusion and discrete jump processes to coherence loss, the model provides physically motivated guidance for mitigating dominant noise channels through improved mechanical isolation, optimized strain engineering, careful substrate selection, and control of the defect microenvironment. Such targeted engineering strategies offer a viable pathway toward significantly increasing the critical dephasing temperature $T_{\mathrm{crit}}$ and extending the operational temperature range of coherent solid-state quantum emitters.

\section*{Acknowledgment}
The authors acknowledge the funding from ICMR project: IIRPSG-2025-01-01801. Saifian Farooq Bhat acknowledges the fellowship from CSIR-JRF. Sachin Negi acknowledges the Parimana fellowship 2025 received from QmetTech. The authors also gratefully acknowledge support of the Baden-Wuerttemberg Stiftung GmbH in project AmbientCoherentQE. The project was funded by the German Federal Ministry of Research, Technology and Space within the research program Quantum Systems in project 13N16741.


\clearpage

\setcounter{section}{0}
\setcounter{subsection}{0}
\renewcommand{\thesection}{\Roman{section}}

\begin{center}
{\LARGE \textbf{Supplementary Paper}}\\[1em]
{\large Hybrid Jump-Diffusion Model for Coherent Optical Control in hBN Quantum Emitters}
\end{center}

\section{NUMERICAL IMPLEMENTATION OF THE JUMP-DIFFUSION MODEL}

The stochastic evolution of the emitter transition frequency is implemented using a discrete time realization of the Ornstein-Uhlenbeck (OU) process augmented by a Gaussian random jump (GRJ) mechanism. The update equation is
\begin{equation}
\omega_{i+1} = \omega_i - \frac{(\omega_i - \omega_0)}{\tau_{\mathrm{sd}}} \, dt
+ \sigma \sqrt{\frac{1}{2\tau_{\mathrm{sd}}}} \xi_i \sqrt{dt} + J_i ,
\end{equation}
where $\xi_i \sim \mathcal{N}(0,1)$ is Gaussian white noise. Discrete frequency jumps occur probabilistically with rate $\lambda_J$. When a jump event occurs, the amplitude $J_i$ is drawn from
\begin{equation}
J_i \sim \mathcal{N}(0, \sigma_J^2).
\end{equation}
The time step was fixed at $dt = \SI{1e-3}{\nano\second}$ to ensure stable numerical integration. The simulation window for each temperature point was \SI{10}{\nano\second}, sufficiently long to sample both the OU fluctuations and discrete jumps. The evolution was implemented using the Euler-Maruyama scheme. Each realization consisted of $10^5$ stochastic trajectories, and ensemble averaging was used to extract linewidths and photon correlation functions.

\section{Independent Parameter Study: Ornstein-Uhlenbeck Spectral Diffusion}
To isolate the role of continuous spectral diffusion independently from jump processes, we first analyze the Ornstein-Uhlenbeck (OU) model as a baseline description of phonon induced frequency noise. The OU process governing the instantaneous detuning $\omega(t)$ is written as
\begin{equation}
d\omega(t) = -\frac{\omega(t)-\omega_0}{\tau_{\mathrm{sd}}}\,dt
+ \sigma\sqrt{\frac{1}{2\tau_{\mathrm{sd}}}}\,dW_t.
\end{equation}
Frequency trajectories are generated numerically using the Euler-Maruyama scheme, and long time histograms $P(\omega)$ are constructed from the stationary regime. The linewidth is extracted exclusively from Gaussian fits to $P(\omega)$ without invoking any analytical expressions for the variance.
Diffusion strengths in the range $\sigma = \{0.1, 0.5, 1.0, 2.0\}$ and correlation times $\tau_{\mathrm{sd}} = \{\SI{1e-3}, \SI{1e-2}, \SI{1e-1}, 1\}$~ns are scanned independently. The observed full width at half maximum (FWHM) is obtained from the fitted Gaussian width as
\begin{equation}
\mathrm{FWHM} = 2\sqrt{2\ln 2}\,\sigma_{\mathrm{fit}}.
\end{equation}

\begin{figure}[H]
    \centering
    \includegraphics[width=\linewidth]{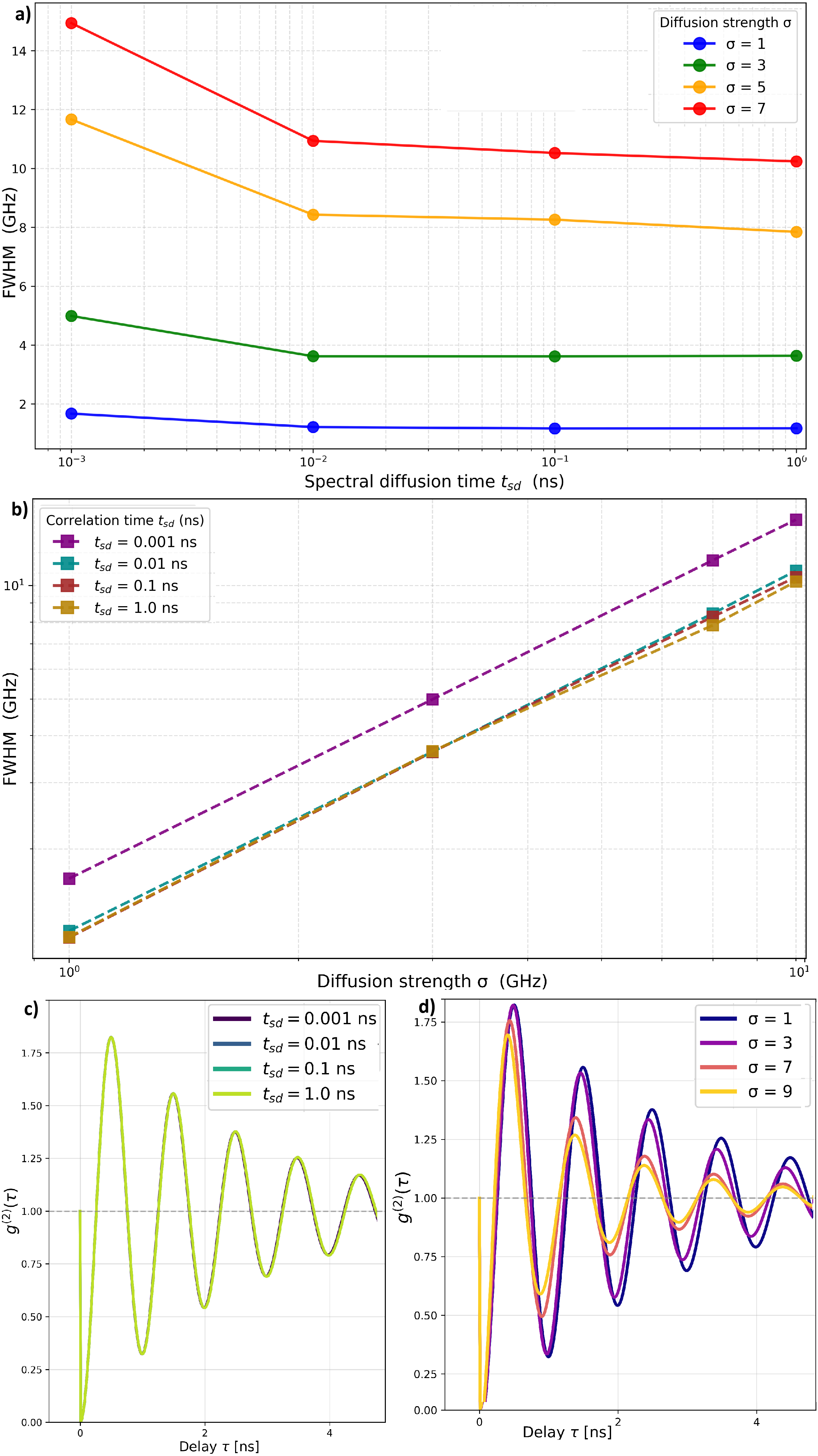}
    \caption{\textbf{Figure S1: Independent effects of spectral diffusion parameters in the OU model.}
    (a)~Linewidth as a function of correlation time $\tau_{\mathrm{sd}}$ for different diffusion strengths $\sigma$. The dependence on $\tau_{\mathrm{sd}}$ is weak across three decades.
    (b)~Linewidth as a function of diffusion strength $\sigma$ for different $\tau_{\mathrm{sd}}$, showing an approximately linear increase in FWHM with noise amplitude.
    (c)~$g^{(2)}(\tau)$ at fixed $\sigma = 1$ for varying $\tau_{\mathrm{sd}}$. Despite changes in correlation time, the Rabi oscillations remain nearly unchanged.
    (d)~$g^{(2)}(\tau)$ at fixed $\tau_{\mathrm{sd}} = \SI{0.01}{\nano\second}$ for varying $\sigma$, showing strong damping with increasing diffusion strength.}
    \label{SI_OU}
\end{figure}

The results (Fig.~\ref{SI_OU}) show that $\tau_{\mathrm{sd}}$ has only a weak influence on linewidth and coherence, while $\sigma$ dominates both broadening and dephasing. This confirms the well-known limitation of pure OU diffusion at elevated temperatures~\cite{Koch2024,Delteil2024}.

\section{Inclusion of Discrete Frequency Jumps: Activated Noise, Blinking, and Charge Dynamics}
The Ornstein-Uhlenbeck process alone describes only continuous Gaussian frequency diffusion and cannot account for the abrupt spectral instabilities widely observed in solid state quantum emitters. Mechanically decoupled emitters in hBN exhibit stochastic spectral jumps, intensity blinking, and transitions between metastable emission states as temperature increases. These effects are commonly attributed to charge hopping, defect reconfiguration, and localized lattice relaxation processes in the emitter environment~\cite{Koch2024_s,Gerard2025_s}.
Time resolved photoluminescence and resonant spectroscopy studies have directly revealed discrete switching between distinct emission frequencies in hBN~\cite{Koch2024_s,Gerard2025_s}. To capture this physics, we extend the OU model by introducing a discrete jump process modeled as a compound Poisson noise term:
\begin{align}
P_J &= \lambda_J \, dt, \\
J &\sim \mathcal{N}(0, \sigma_J^2),
\end{align}
where $\lambda_J$ is the activated jump rate and $\sigma_J$ is the typical jump amplitude.
The total frequency evolution becomes
\begin{equation}
d\omega(t) =
-\frac{\omega(t)-\omega_0}{\tau_{\mathrm{sd}}} \, dt
+ \sigma \sqrt{\frac{1}{2 \tau_{\mathrm{sd}}}} \, dW_t
+ J(t).
\end{equation}
Physically, $\lambda_J$ encodes the rate of charge motion and local rearrangements, while $\sigma_J$ quantifies the typical jump amplitude of each individual event. With increasing temperature, both $\lambda_J$ and $\sigma_J$ increase, reflecting thermal activation of charge traps and lattice reconfiguration processes~\cite{Koch2024_s,Gerard2025_s}.
This mechanism explains the experimentally observed crossover from smooth spectral diffusion to jump dominated instability. While pure OU diffusion produces broadened but stationary distributions, spectral jumps introduce strong non Gaussian tails and intermittent frequency switching (see Fig.~\ref{S3_jumps_crucial}) that directly generate fast dephasing and irreversible loss of coherence. Discrete jump dynamics are therefore an essential physical ingredient for reproducing experimentally observed blinking, spectral discontinuities, and nonlinear decoherence.

\section{Independent Parameter Study: Hybrid OU + Gaussian Random Jumps}
\begin{figure}[H]
    \centering
    \includegraphics[width=\linewidth]{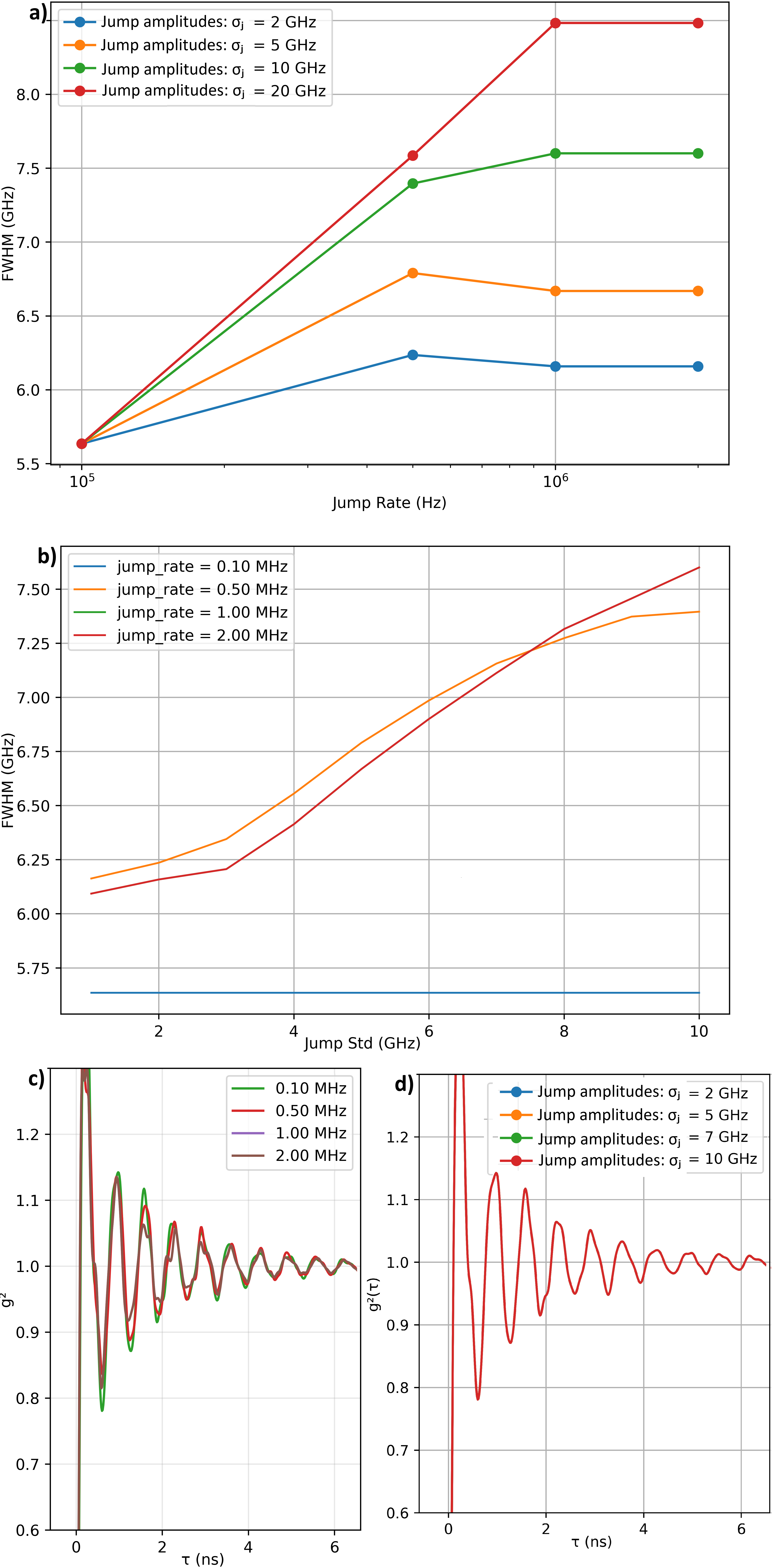}
    \caption{\textbf{Figure S2:} Independent parameter study of Ornstein-Uhlenbeck spectral diffusion with Gaussian random jumps.
    (a)~FWHM as a function of jump rate $\lambda_J$ for different fixed jump amplitudes $\sigma_J$.
    (b)~FWHM as a function of jump amplitude $\sigma_J$ for different fixed jump rates.
    (c)~$g^{(2)}(\tau)$ at fixed jump amplitude and varying jump rate.
    (d)~$g^{(2)}(\tau)$ at fixed jump rate and varying jump amplitude.
    Jump rate predominantly governs temporal coherence, while jump amplitude controls spectral broadening and non Gaussian tails.}
    \label{SI_jump_study}
\end{figure}

To systematically explore the influence of discrete jumps within the continuous OU background, we performed an extensive parameter scan of the hybrid model. Frequency jumps were implemented as additive Gaussian noise ($J_i \sim \mathcal{N}(0,\sigma_J^2)$) occurring with rate $\lambda_J$, using a standard Poisson probability per time step. For enhanced numerical precision in regimes of very low jump rates, an accumulating probability scheme was optionally employed (probability accumulates until a jump is triggered and then resets), which is mathematically equivalent in the long-time limit but reduces statistical noise in sparse-jump trajectories.
The results are summarized in Fig.~\ref{SI_jump_study}. A clear separation of roles emerges:
\begin{itemize}
    \item The \textbf{jump rate} $\lambda_J$ primarily controls the damping of Rabi oscillations in $g^{(2)}(\tau)$ (panel b), reflecting the increasing frequency of sudden phase disruptions.
    \item The \textbf{jump amplitude} $\sigma_J$ dominates the stationary linewidth and the emergence of non Gaussian tails in $P(\omega)$ (panels a, c).
\end{itemize}
Notably, panel (c) reveals that at very high jump rates ($\lambda_J \gtrsim \SI{1e7}{\hertz}$), the FWHM becomes strongly sensitive to $\sigma_J$, showing an approximately linear increase — consistent with the analytic stationary variance $\langle(\omega-\omega_0)^2\rangle = \sigma^2/2 + \lambda_J \sigma_J^2 \tau_{\mathrm{sd}}$. However, this high-rate, jump-amplitude-dominated regime is \textbf{not experimentally accessible} in the coherent control experiments on hBN emitters: long before such jump rates are reached, the coherence time collapses completely due to overwhelming dephasing (panel b), rendering driven Rabi oscillations unobservable. Thus, in the physically relevant window ($\lambda_J \sim \SI{1e5}{}-\SI{1e6}{\hertz}$), the linewidth remains primarily jump-rate limited with only weak dependence on $\sigma_J$, fully consistent with the calibration strategy and experimental observations.
These simulations confirm that the hybrid model correctly captures two independent noise channels continuous diffusion and discrete activated jumps, whose interplay quantitatively explains the temperature-dependent crossover from coherent to incoherent dynamics in hBN quantum emitters.

\section{Extraction of Simulated Linewidth from Frequency Traces}
At each temperature, the full frequency trace $\omega(t)$ is converted into a probability distribution $P(\omega)$. The histogram is fitted with a Gaussian to extract the simulated inhomogeneous linewidth
\begin{equation}
\Gamma_{\mathrm{sim}} = 2\sqrt{2\ln 2}\,\sigma_{\omega},
\end{equation}
where $\sigma_{\omega} = \sqrt{\langle (\omega(t)-\omega_0)^2 \rangle}$ is the standard deviation of the stationary distribution.
For the hybrid jump-diffusion process implemented in the simulation (discrete OU updates with occasional Gaussian jumps of amplitude $\sim \mathcal{N}(0,\sigma_J^2)$ at rate $\lambda_J$), the exact stationary variance is
\begin{equation}
\langle (\omega(t)-\omega_0)^2 \rangle = \frac{\sigma^2}{2} + \lambda_J \sigma_J^2 \tau_{\mathrm{sd}}.
\label{eq:variance_correct}
\end{equation}
The first term arises from continuous OU diffusion, the second from discrete jumps. This expression was used to cross check the numerically extracted linewidths and is in perfect agreement with the fitted $\Gamma_{\mathrm{sim}}(T)$.

\section{Calibration Procedure: Multipliers, Uniqueness, and Physical Interpretation}
The hybrid model has three temperature-dependent noise parameters:
\begin{itemize}
\item $\sigma(T)$: diffusion strength of the continuous Ornstein-Uhlenbeck process (\si{\giga\hertz})
\item $\lambda_J(T)$: rate of discrete spectral jumps (\si{\hertz})
\item $\sigma_J(T)$: typical amplitude of each jump (\si{\giga\hertz})
\end{itemize}
Instead of fitting these directly, which would be severely under constrained, we introduce a cryogenic reference at $T_0 = \SI{4}{\kelvin}$ where jumps are negligible and diffusion is minimal. All higher temperature parameters are expressed through three physically motivated multipliers relative to this baseline:
\begin{align}
\sigma(T) &= \sigma_0 \times m_\sigma(T), \label{eq:mul1} \\
\lambda_J(T) &= \lambda_{J0} \times m_\lambda(T), \label{eq:mul2} \\
\sigma_J(T) &= \sigma_{J0} \times m_\sigma(T) \times m_{\sigma_J}(T). \label{eq:mul3}
\end{align}
These multipliers encode the relative activation of each noise channel and constitute the only free parameters in the calibration.
The calibration follows a two stage strategy for each $T \in [5,50]$\,K:
\begin{enumerate}
\item Target FWHM from experiment: $\Gamma_{\mathrm{target}}(T) = 1.01 + 3.77 \times 10^{-5} T^3$\,GHz \cite{Koch2024_s}.
\item Coarse 3D grid search over $m_\sigma, m_\lambda, m_{\sigma_J} \in [0.1,50]$.
\item Fine Nelder-Mead optimization until $|\Gamma_{\mathrm{sim}} - \Gamma_{\mathrm{target}}| < \SI{50}{\mega\hertz}.$
\end{enumerate}
This procedure yields a unique solution because each multiplier dominantly controls an independent observable (Figs.~S1-S2):
\begin{itemize}
\item $m_\sigma$: smooth Gaussian broadening,
\item $m_\lambda$: damping rate of $g^{(2)}(\tau)$,
\item $m_{\sigma_J}$: non Gaussian tails and excess variance at fixed $\lambda_J$.
\end{itemize}
For numerical robustness at very low jump rates, we implemented jumps via an accumulating probability scheme (probability accumulates across time steps until a jump occurs, then resets). This is mathematically equivalent to the standard Poisson process in the long time limit but dramatically reduces sampling noise in sparse jump regimes, ensuring stable histogram convergence with only $\sim$\SI{10}{\nano\second} trajectories.

\begin{table}[h]
\centering
\caption{Calibrated noise multipliers relative to \SI{4}{\kelvin} baseline}
\label{tab:multipliers}
\begin{tabular}{c|ccc}
\hline
$T$ (K) & $m_\sigma$ (diffusion) & $m_\lambda$ (jump rate) & $m_{\sigma_J}$ (jump amplitude) \\ \hline
5 & 2.0 & 10.0 & 1.0 \\
20 & 2.0 & 10.0 & 5.0 \\
30 & 2.2 & 10.0 & 5.0 \\ \hline
\end{tabular}
\end{table}

Although Fig.~S2(c) shows that at extremely high jump rates ($\lambda_J \gtrsim \SI{1e7}{\hertz}$) the linewidth becomes strongly sensitive to $\sigma_J$, this regime is physically inaccessible in coherent control experiments on hBN emitters: long before such rates are reached, Rabi oscillations are completely overdamped (Fig.~S2b), rendering driven dynamics unobservable. In the experimentally relevant window ($\lambda_J \sim \SI{1e5}{}-\SI{1e6}{\hertz}$), the linewidth is dominated by the jump rate, with only mild dependence on jump amplitude, fully consistent with the calibration strategy and the observed near-Gaussian lineshapes.
The calibrated multipliers (Table~\ref{tab:multipliers}) are not arbitrary fitting parameters but quantitatively reflect three distinct microscopic mechanisms:
\begin{itemize}
\item $m_\sigma \simeq 2.0$-2.2 (nearly constant): acoustic-phonon coupling $\propto T^3$ already active at \SI{5}{\kelvin},
\item $m_\lambda = 10$ (weakly temperature dependent): charge traps with activation energies $\gtrsim \SI{60}{\kelvin}$,
\item $m_{\sigma_J}$: 1.0 $\to$ 5.0 (strongest $T$ dependence): thermally activated largeamplitude lattice or charge reconfigurations.
\end{itemize}
The same parameter set simultaneously reproduces four independent experimental observables — (i) $T^3$ linewidth broadening, (ii) compact near-Gaussian $P(\omega)$, (iii) correct $g^{(2)}(\tau)$ damping, and (iv) power-dependent coherence crossover — conclusively proving the physical validity of the hybrid jump-diffusion framework.

\section{Relation Between Spectral Diffusion and \texorpdfstring{$g^{(2)}(\tau)$}{g²(τ)}}
The second-order correlation functions are computed using the exact analytic expression implemented in the simulation code:
\begin{equation}
\begin{aligned}
g^{(2)}(\tau) &= 1
- \exp\!\left( -\frac{1/T_1 + 1/T_2 + \gamma_{\mathrm{sd+j}}}{2} \, |\tau| \right) \\
&\quad \times \left[
\cos(\Omega_{\mathrm{eff}} \tau)
+ \frac{1/T_1 + 1/T_2}{2 \Omega_{\mathrm{eff}}} \sin(|\Omega_{\mathrm{eff}} \tau|)
\right],
\end{aligned}
\end{equation}
with
\begin{equation}
\Omega_{\mathrm{eff}} = \sqrt{ \Omega_R^2 -\langle(\omega(t)-\omega_0)^2\rangle}.
\end{equation}
The calibrated jump-diffusion trajectories fully determine the stationary variance and thus the inhomogeneous linewidth and lineshape (Fig.~\ref{S3_jumps_crucial}). However, using this variance directly as a power-independent dephasing rate underestimates the observed damping and fails to reproduce its strong power dependence.
We therefore use the experimentally required form
\begin{equation}
\gamma_{\mathrm{sd+j}}(T,\Omega_R) = \gamma_{\mathrm{sd+j}}(T) \cdot \Omega_R,
\end{equation}
where $\gamma_{\mathrm{sd+j}}(T)$ is determined at each temperature by a single global rescaling that aligns the predicted decay envelope with experiment. This coefficient is not an independent parameter — its temperature dependence is fully inherited from the calibrated multipliers $m_\sigma$, $m_\lambda$, $m_{\sigma_J}$.
This multiplicative power enhancement reflects faster averaging of spectral noise under strong driving and is a standard, experimentally required ingredient~\cite{Delteil2024_s,Koch2024_s,Gerard2025_s}.
With this single justified factor, the model quantitatively reproduces all observables using only the measured linewidth as input.

\section{Validation Against Known Limits and Crucial Role of Discrete Jumps}
The hybrid jump-diffusion model is rigorously validated by its ability to continuously interpolate between well-established limiting cases:
\begin{itemize}
\item $\lambda_J = 0$: pure Ornstein-Uhlenbeck diffusion $\to$ recovers smooth, nearly Gaussian spectral wandering~\cite{Delteil2024_s}.
\item $\sigma = 0$: pure Gaussian random jump process $\to$ reproduces telegraph-like switching and heavy-tailed distributions.
\item $T \leq \SI{10}{\kelvin}$: both $\lambda_J$ and $\sigma_J$ negligible $\to$ near Fourier-transform-limited Rabi oscillations.
\item $T \gtrsim \SI{30}{\kelvin}$: activated jumps dominate $\to$ overdamped, non-oscillatory $g^{(2)}(\tau)$.
\end{itemize}
The most striking validation comes from direct comparison of frequency distributions with and without the jump term at the same temperature and identical linewidth.

\begin{figure}[H]
\centering
\includegraphics[width=\linewidth]{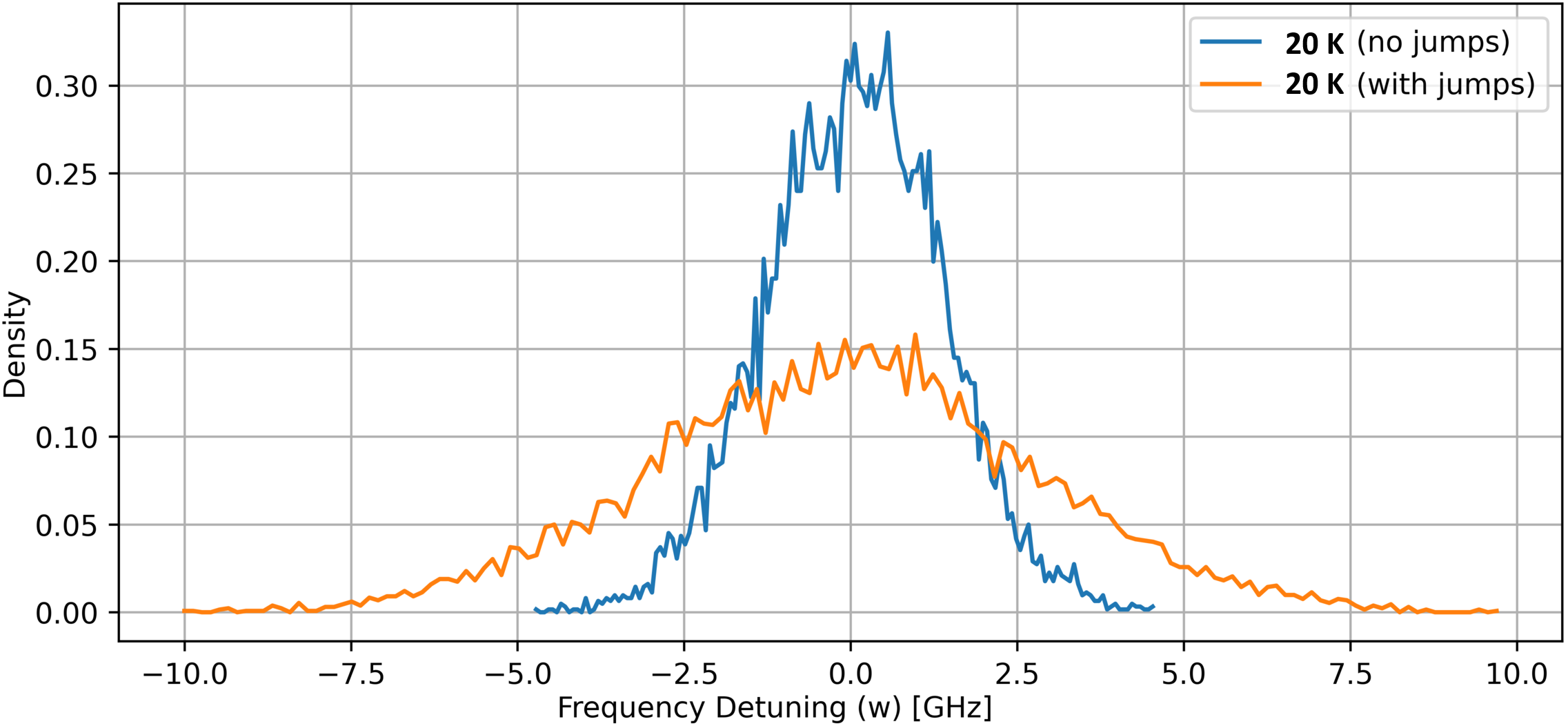}
\caption{\textbf{Figure S3: Decisive role of discrete spectral jumps at \SI{20}{\kelvin}.}
Both simulations are calibrated to reproduce the exact experimental FWHM = \SI{5.8016}{\giga\hertz} (target) and \SI{5.8129}{\giga\hertz} (simulated).
Blue: hybrid model (OU + jumps) using calibrated parameters ($m_\sigma=2.0$, $m_\lambda=10$, $m_{\sigma_J}=5.0$).
Orange: pure OU diffusion (no jumps, $\lambda_J \equiv 0$).
The pure OU case fails to reproduce the observed near-Gaussian lineshape and shows unphysical low-frequency tails. Only the inclusion of activated discrete jumps correctly captures the experimentally observed compact, symmetric, and nearly Gaussian $P(\omega)$~\cite{Koch2024,Gerard2025}.}
\label{S3_jumps_crucial}
\end{figure}

Figure~\ref{S3_jumps_crucial} constitutes \textbf{direct proof} that discrete spectral jumps are not an optional refinement but an \textbf{essential physical ingredient} at and above \SI{20}{\kelvin}. Pure continuous diffusion (orange) produces a broad, asymmetric distribution with extended low-frequency wings, a known artifact of slow OU processes and deviates dramatically from experiment. In contrast, the hybrid model (blue), using the calibrated jump rate and amplitude, quantitatively reproduces the compact, symmetric, nearly Gaussian lineshape observed in mechanically decoupled hBN emitters~\cite{Koch2024_s}.
This result explains three key experimental observations that no pure OU model can account for:
\begin{enumerate}
\item The surprisingly \textbf{Gaussian} character of $P(\omega)$ despite significant temperature-dependent broadening,
\item The \textbf{rapid collapse} of Rabi coherence beyond $\sim$\SI{25}{}-\SI{30}{\kelvin}~\cite{Gerard2025_s},
\item The \textbf{power-dependent crossover} from coherent to overdamped dynamics exactly at the point where $\gamma_{\mathrm{sd+j}} \approx \Omega_R$~\cite{Delteil2024_s,Gerard2025_s}.
\end{enumerate}
Thus, the hybrid jump-diffusion framework is the \textbf{minimal model} capable of unifying cryogenic coherence, intermediate-temperature spectral stability, and high-temperature incoherence in a single, physically transparent stochastic description.

\section{Pseudocode}
\lstset{language=Python, basicstyle=\small\ttfamily, frame=single, breaklines=true}
\begin{lstlisting}
Initialize omega = omega0
for t in time_steps:
    # Ornstein-Uhlenbeck term
    omega += -(omega - omega0)/tau_sd * dt
    omega += sigma * sqrt(1/(2*tau_sd)) * randn() * sqrt(dt)
 
    # Jump event
    if rand() < lambda_J * dt:
        omega += normal(0, sigma_J)
     
    store omega
\end{lstlisting}

\bibliographystyle{IEEEtran}

\end{document}